\documentclass[conference]{IEEEtran}
\IEEEoverridecommandlockouts
\usepackage{cite}
\usepackage{amsmath,amssymb,amsfonts}
\usepackage{algorithmic}
\usepackage{graphicx}
\usepackage{textcomp}
\usepackage{xcolor}
\usepackage{multirow}

\usepackage{tikz}
\usepackage{textcomp}
\usepackage{hyperref}
\usepackage{lipsum}

\newcommand\copyrighttext{%
  \footnotesize \textcopyright 2025 Personal use of this material is permitted. Permission from IEEE must be obtained for all other uses, including reprinting/republishing this material for advertising or promotional purposes, collecting new collected works for resale or redistribution to servers or lists, or reuse of any copyrighted component of this work in other works.
  }
\newcommand\copyrightnotice{%
\begin{tikzpicture}[remember picture,overlay]
\node[anchor=south,yshift=10pt] at (current page.south) {\fbox{\parbox{\dimexpr\textwidth-\fboxsep-\fboxrule\relax}{\copyrighttext}}};
\end{tikzpicture}%
}

\def\BibTeX{{\rm B\kern-.05em{\sc i\kern-.025em b}\kern-.08em
    T\kern-.1667em\lower.7ex\hbox{E}\kern-.125emX}}
\begin{document}

\title{Multiparty Selective Disclosure using Attribute-Based Encryption}

\author{\IEEEauthorblockN{Shigenori Ohashi}
\IEEEauthorblockA{\textit{NTT Social Informatics Laboratories} \\
\textit{}\\
 \\
}
}

\maketitle
\copyrightnotice

\begin{abstract}
This study proposes a mechanism for encrypting SD-JWT (Selective Disclosure JSON Web Token) Disclosures using Attribute-Based Encryption (ABE) to enable flexible access control on the basis of the Verifier's attributes. By integrating Ciphertext-Policy ABE (CP-ABE) into the existing SD-JWT framework, the Holder can assign decryption policies to Disclosures, ensuring information is selectively disclosed. The mechanism's feasibility was evaluated in a virtualized environment by measuring the processing times for SD-JWT generation, encryption, and decryption with varying Disclosure counts (5, 10, 20). Results showed that SD-JWT generation is lightweight, while encryption and decryption times increase linearly with the number of Disclosures. This approach is suitable for privacy-sensitive applications like healthcare, finance, and supply chain tracking but requires optimization for real-time use cases such as IoT. Future research should focus on improving ABE efficiency and addressing scalability challenges.
\end{abstract}

\begin{IEEEkeywords}
Verifiable credentials, VCs, Attribute-based encryption, ABE, Selective disclosure, SD-JWT
\end{IEEEkeywords}

\section{Introduction}
The importance of digital identity management has grown significantly with the rapid digitalization of online transactions and service provision. In particular, protecting privacy while selectively sharing necessary information has become a critical challenge. Conventional digital identity systems often share users’ personal information in its entirety, raising concerns about privacy breaches and security risks due to excessive data sharing. Against this backdrop, there is a growing demand for technologies that allow users to control their own information and share it within appropriate boundaries\cite{b25}.

One such technology designed to address this need is Selective Disclosure JSON Web Token (SD-JWT)\cite{b2}. SD-JWT uses digital signatures to enable users to selectively disclose only the information necessary to Verifiers. This allows users to securely share minimal information with Verifiers. However, the current implementation of SD-JWT has several limitations. A key issue is that the disclosed information (Disclosure) is static, making it challenging to provide different information dynamically to different Verifiers. In its current form, the same Disclosure is shared with all Verifiers, making it impossible to tailor information sharing to Verifiers with varying roles or permissions. This limitation reduces flexibility, particularly in use cases involving multiple stakeholders, and poses challenges from a privacy protection perspective.

Attribute-Based Encryption (ABE)\cite{b3}, on the other hand, is a cryptographic technique that enables access control by associating conditions, known as "attributes," with encrypted data. ABE allows only users who meet specific conditions to decrypt the ciphertext, thereby enabling flexible access control. Specifically, Ciphertext-Policy Attribute-Based Encryption (CP-ABE) allows encryption policies to be set during the encryption process, and decryption is permitted only if the attributes of the decryption key satisfy the defined policy. By leveraging this feature, the flexibility of selective information sharing and access control can be enhanced.

This study proposes a novel mechanism that combines SD-JWT with ABE to enable information to be selectively disclosed on the basis of the attributes of Verifiers. Specifically, the Disclosure component of SD-JWT is encrypted using ABE, and decryption is controlled on the basis of the attribute information of the Verifiers. This approach allows information to be tailored to Verifiers with different roles or permissions, minimizes the risk of data leakage, and effectively protects user privacy.

The proposed approach offers three key features:
\begin{itemize}
    \item Flexible information disclosure control: The content of Disclosure can be dynamically adjusted on the basis of the attributes of Verifiers, significantly enhancing the flexibility of information sharing.
    \item Enhanced privacy protection: By sharing only the minimum necessary information, the approach prevents information disclosure beyond a Verifier's permissions and safeguards user privacy.
    \item Security and efficiency: By combining the strengths of SD-JWT and ABE, the proposed mechanism achieves high security while maintaining efficient encryption and decryption processes.
\end{itemize}

The proposed system can be applied to a wide range of use cases in digital identity management. For example, it can be utilized in the sharing of patient information in the healthcare domain, identity verification (KYC: Know Your Customer) processes in financial institutions, and supply chain tracking. In these fields, a mechanism is needed that enables appropriate information to be shared among multiple stakeholders while protecting user privacy.
This paper details the design of the proposed mechanism. Additionally, the feasibility of the overall system is evaluated. 

The paper is organized as follows: Section 2 discusses related research, Section 3 describes the proposed method and valid use cases, Section 4 evaluates the proposed method and discusses the results, and Section 5 summarizes.

\section{Related work}
\subsection{Verifiable Credentials}
As challenges in digital identity management become increasingly prominent\cite{b25}, Verifiable Credential (VC) has emerged as a technology that balances privacy protection and information reliability. VC provides a framework that allows users to manage and share their qualifications and attribute information digitally and verifiably. Its roots can be traced back to the late 1990s, when Public Key Infrastructure (PKI) and digital certificates began to gain prominence. PKI ensures the reliability and authentication of data through digital signatures and encryption, but its centralized model has been criticized for issues such as privacy breaches and single points of failure\cite{b22}.

Against this backdrop, the World Wide Web Consortium (W3C) proposed the "Verifiable Credentials Data Model 1.0" in 2017, marking a significant step in standardizing the VC framework\cite{b1}. This model integrates with Decentralized Identity (DID), enabling users to retain control over their information. This approach goes beyond traditional centralized systems and introduces a flexible and privacy-preserving decentralized model.

The use cases proposed by W3C demonstrate the practicality and wide applicability of VCs\cite{b23}. For instance, government-issued digital IDs can streamline online service registration and identity verification. Similarly, educational credentials such as degree certificates and qualifications can be digitally issued and shared efficiently. In the healthcare sector, patients can share only the necessary information with specific healthcare providers, thereby protecting their privacy while improving service efficiency. In the financial sector, VCs can optimize KYC processes in banks and financial institutions, reducing risks associated with fraud and data breaches. For international travel, digital passports and visas can expedite airport and immigration procedures. Additionally, e-commerce applications often involve the purchase of age-restricted products, where VCs enable users to share only minimal information to prove their eligibility.

VCs, owing to their flexibility and versatility, have led to the development of various specifications. For instance, Selective Disclosure JSON Web Token (SD-JWT) is an extension of JSON Web Token (JWT) designed to enable selective disclosure\cite{b2}. An SD-JWT issued by an Issuer contains a core token and Disclosures. The Holder can share only the required Disclosures with a specific Verifier, who can then validate the token’s authenticity by using the Issuer’s signature. This mechanism is well-suited for use cases in domains such as healthcare and finance, where the risk of data leakage needs to be minimized while enabling information to be flexibly shared.

Similarly, Anonymous Credentials (AnonCreds), a specification adopted by the Hyperledger Indy project, leverages Zero-Knowledge Proofs (ZKP) to achieve advanced anonymity\cite{b24}. Holders can share only the necessary information with Verifiers while keeping other details undisclosed. This feature is particularly valuable in scenarios where strong privacy protections are required \cite{b4}.

Additionally, Linked Data Proof Verifiable Credentials (LDP-VC), proposed by W3C, use Linked Data Proofs and adopt the JSON-LD format\cite{b5}. This specification is advantageous for semantic data processing and interoperability\cite{b26}. It ensures the trustworthiness of VCs through digital signatures and facilitates flexible identity management in decentralized environments when integrated with DIDs.

The widespread adoption of VCs necessitates robust privacy protection for users, with unlinkability being a critical requirement. Unlinkability refers to the property that prevents the correlation of VCs used across different services or transactions by the same user. Techniques leveraging ZKP enable Holders to prove necessary information without revealing unique identifiers \cite{b4}. Policy-based encryption that combines VCs with ABE is also being researched to prevent linking across transactions.

\subsection{Attribute-Based Encryption}
ABE is a type of public-key encryption designed to enable flexible access control\cite{b19}. Unlike traditional public-key encryption, where encryption and decryption are tied to specific individuals or key pairs, ABE allows access to encrypted data to be governed by "attributes" or "policies." This feature makes ABE particularly beneficial in scenarios that require secure information sharing and privacy protection.

The concept of ABE was first introduced in 2005 by Sahai and Waters in their research on "Fuzzy Identity-Based Encryption." This model extended Identity-Based Encryption (IBE) by using attributes as identifiers and allowed decryption if the attributes embedded in the ciphertext matched a subset of the user's attributes\cite{b3}. Subsequently, in 2006, Goyal, Pandey, Sahai and Waters proposed one of major models of ABE: Key-Policy Attribute-Based Encryption (KP-ABE)\cite{b9}. Then in 2007, Bethencourt, Sahai and Waters proposed Ciphertext-Policy Attribute-Based Encryption (CP-ABE)\cite{b10}. In CP-ABE, access policies are embedded in the ciphertext, and decryption is possible if the user's decryption key satisfies the policy. Conversely, KP-ABE incorporates the access policy into the decryption key, allowing decryption if the attributes of the ciphertext match the policy encoded in the key. These two models significantly expanded the applicability of ABE, particularly in addressing access control challenges in cloud computing and decentralized systems.

ABE has given rise to many derivative models, each tailored to different applications. For instance, Hierarchical Attribute-Based Encryption (HABE) supports multi-level access control in organizations or systems \cite{b12}. Dynamic ABE allows attributes or policies to be updated dynamically, making it suitable for systems requiring real-time access control\cite{b20}. Additionally, Large Universe ABE expands the attribute space, improving scalability and flexibility \cite{b13}

One of the defining features of ABE is its ability to enable flexible access control. By integrating access policies into the encryption process, data providers can implement fine-grained access control \cite{b9}. For example, access control can be based on specific job roles, geographic locations, or organizational hierarchies. Furthermore, ABE is highly effective in preserving privacy. Encrypted data can only be decrypted by users who meet the specified conditions, preventing unnecessary data leakage while allowing essential information sharing \cite{b10}. This makes ABE particularly valuable in domains such as healthcare and finance, where robust privacy protection is critical.

In addition to its privacy-preserving capabilities, ABE is well-suited for decentralized environments. Since it does not require a central authority, ABE can function effectively in distributed platforms like cloud storage or blockchain systems \cite{b11}. Moreover, ABE can be integrated with ZKP to enhance privacy further. This combination enables users to prove their decryption capabilities without disclosing actual data, thereby achieving even stronger privacy guarantees \cite{b21}.


\section{Proposed method}
\subsection{Details of our idea}
\begin{figure*}
    \centering
    \includegraphics[width=1.0\linewidth]{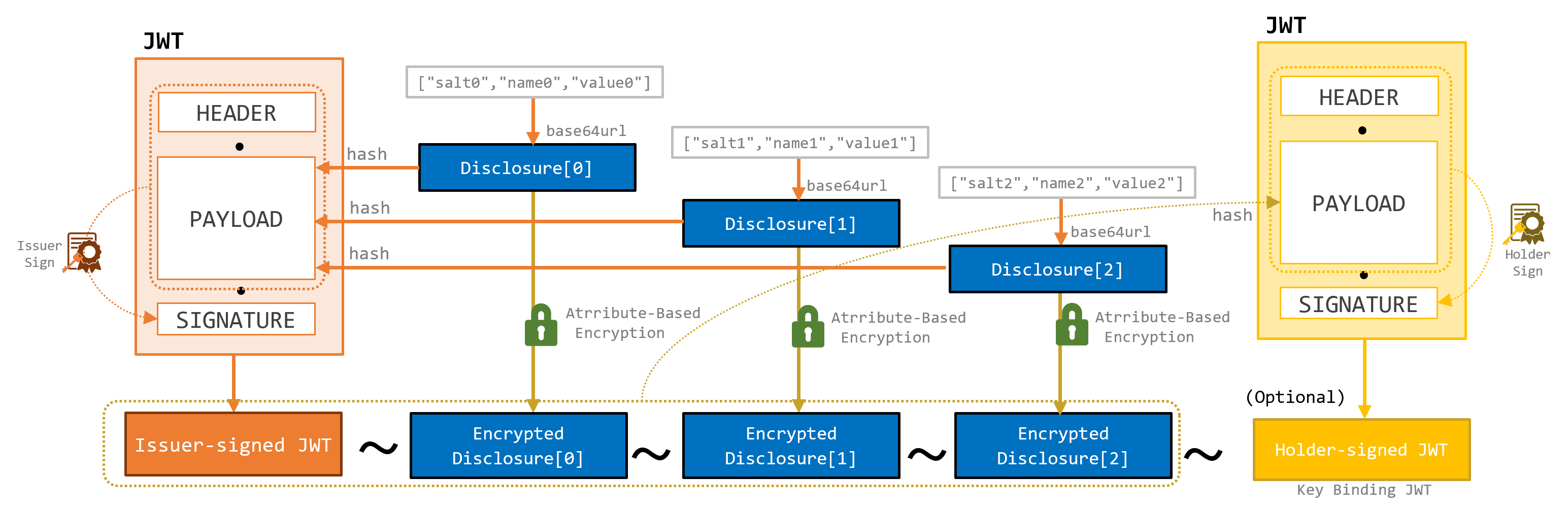}
    \caption{Overall picture of SD-JWT with attribute-based encryption}
    \label{fig1}
\end{figure*}

This section explains in detail a mechanism for encrypting SD-JWT Disclosures using ABE, allowing information to be selectively disclosed on the basis of the attributes of the Verifier (Figure\ref{fig1}). This approach extends the flexibility of existing SD-JWT selective disclosure features by integrating ABE to enable more granular privacy protection and flexible access control. The mechanism involves three key roles: the Issuer, Holder, and Verifier. The Issuer issues a basic SD-JWT to the Holder. The Holder encrypts the Disclosures contained in the SD-JWT using CP-ABE and attaches specific decryption conditions. Subsequently, the Holder generates a Key Binding JWT and provides both the SD-JWT and Key Binding JWT to the Verifier. The Verifier verifies the JWTs signed by the Issuer and Holder and retrieves the Disclosures they are authorized to access on the basis of their attributes.

The Issuer issues a standard SD-JWT to the Holder. This SD-JWT consists of an Issuer-signed JWT concatenated with a series of Disclosures. The Issuer-signed JWT includes the hash values of all the Disclosures in its payload to ensure their integrity. 
The Issuer signs the JWT, ensuring its authenticity and preventing tampering. The Disclosures themselves are provided to the Holder, who performs the subsequent encryption process.

The Holder encrypts the Disclosures contained in the SD-JWT using CP-ABE. This encryption process allows each Disclosure to be protected with specific decryption conditions. For example, one Disclosure might include the condition "Decryptable if the Verifier possesses attributes A and B," while another Disclosure might include "Decryptable if the Verifier possesses attributes C or D." These conditions are used to strictly manage access control. This process ensures that Disclosures remain inaccessible to Verifiers who do not meet the decryption conditions.

The Holder generates a Key Binding JWT containing the hash value of the encrypted Disclosures and the Issuer-signed JWT. This JWT plays a crucial role in allowing the Verifier to verify that the received Disclosures have not been tampered with. The payload of the Key Binding JWT is structured as a table \ref{tab1}. The payload has a new field to describe the decryption policy for encrypted Disclosures to improve decryption efficiency. The Holder signs the Key Binding JWT to prevent its tampering and provides both the SD-JWT and the Key Binding JWT to the Verifier for further validation. 

\begin{table*}[hbt]
    \centering
    \begin{tabular}{|c|c|c|p{11cm}|}
      \hline
      Item types   & Item names & Necessity & Description\\
      \hline
      \multirow{2}{*}{Header}  & alg  & Must & As required by the JWT specification (RFC 7519) \\
                               & typ  & Must & "kb+abe+jwt" (Newly defined)\\
      \hline
      \multirow{6}{*}{Payload} & iat & Must & Issue date \\
                               & aud & Must & Verifiers of Key Binding JWT \\
                               & nonce & Must & String for replay attack mitigation\\
                               & sd\_hash & Must & Hash value of the Issuer-signed JWT and the encrypted Disclosures \\
                               & \_sd (Newly defined)& Option & An array of hashes of the data before encryption of the Disclosure. The order and number of elements must match the order and number of elements of the encrypted disclosure.\\
                               & abe\_pcy (Newly defined)& Option & An array of decryption policies for encrypted disclosures. The order and number of elements must match the order and number of elements of the encrypted disclosure.\\\hline
    \end{tabular}
    \caption{Key Binding JWT}
    \label{tab1}
\end{table*}

The Verifier performs validation and decryption using the Issuer-signed JWT and Key Binding JWT provided by the Holder. The process involves several steps to ensure the authenticity, integrity, and access control of the Disclosures. First, the Verifier validates the signature of the Issuer-signed JWT using the Issuer’s public key to determine whether it is a legitimate JWT issued by the Issuer. Next, the Verifier validates the signature of the Key Binding JWT by using the Holder's public key, which is referenced in the cnf (Confirmation Key) field of the Issuer-signed JWT. This step determines whether the Key Binding JWT was legitimately generated by the Holder.

The Verifier then ensures that the hash value of the encrypted Disclosures and Issuer-signed JWT in the Key Binding JWT match the received encrypted Disclosures and Issuer-signed JWT. This step guarantees that the encrypted Disclosures have not been tampered with. Subsequently, the Verifier decrypts the encrypted Disclosures using their private key if they possess the required attributes. If decryption is successful, the Verifier computes the hash of the decrypted Disclosure and verifies that it matches the corresponding hash value in the Issuer-signed JWT’s payload. This step determines whether the decrypted Disclosure matches the original Disclosure issued by the Issuer.

Finally, the Verifier retrieves the decrypted Disclosure if they meet the decryption conditions. This information can then be used to complete necessary actions or evaluations. If the Verifier does not meet the decryption conditions, they are unable to access the Disclosure, thereby preventing unauthorized access and preserving data privacy.

This mechanism provides several advantages, including flexible access control, tamper resistance, and robust privacy protection. Information is controlled on the basis of the Verifier's attributes, reducing the risk of unnecessary information sharing or privacy breaches. The use of Issuer-signed JWTs and Key Binding JWTs ensures the integrity and reliability of the shared data.

\subsection{Use Cases}
The combination of SD-JWT and ABE offers a powerful framework for securely sharing and controlling sensitive information on the basis of the attributes of the Verifier. This approach ensures that only authorized Verifiers with the required attributes can decrypt specific portions of the data, enabling fine-grained and privacy-preserving access control. This section proposes three applicable use cases for this framework, providing detailed examples and references to support its effectiveness.

\subsubsection{Use Case 1: Privacy-Preserving Healthcare Data Sharing}
In community healthcare, various stakeholders—including hospitals, clinics, pharmacies, caregivers, visiting nurses, and local volunteer supporters—need to share patient-related information such as medical records, prescriptions, care plans, and health management data. However, traditional medical information-sharing systems rely on centralized databases maintained by hospitals and insurance providers, posing risks such as unauthorized access, privacy violations, and excessive data sharing. Additionally, in many cases, patient information is shared too broadly, raising concerns about privacy protection. By integrating SD-JWT and ABE, our proposed system enables medical and caregiving information to be securely and selectively disclosed in community healthcare settings.

In this use case, an Issuer (e.g., hospitals, clinics, or public health institutions) issues an SD-JWT containing a patient’s medical records, prescription data, and care plans. The Holder (i.e., the patient or their family) encrypts specific medical data using ABE, defining decryption policies on the basis of attributes such as “Hospitals,” “Pharmacies,” “Caregivers,” “Visiting Nurses,” and “Local Supporters.” For instance, the hospital may receive full access to medical records, the pharmacy may only access prescription-related data, and caregivers or visiting nurses may receive information about rehabilitation plans and medication schedules. Verifiers (the various stakeholders) can only decrypt and access information permitted by the patient, ensuring that each entity receives only the necessary data for their role. This mechanism allows patients to precisely control their healthcare information while ensuring appropriate support.

Additionally, this system contributes to integrated community care, which is essential for enabling elderly individuals to live safely within their communities. Information needs to be effectively shared between not only hospitals and care facilities but also pharmacies and local supporters (such as volunteers, municipal workers, and neighbors). For example, if a patient is receiving in-home care, their visiting nurse or caregiver can access only essential health management data, while community supporters receive limited details such as emergency contact information and daily assistance requirements. This enables a collaborative healthcare and welfare network while safeguarding patient privacy.

\subsubsection{Use Case 2: Financial Services and Business Loans}
When a company applies for a business loan, multiple entities—such as financial institutions, credit guarantee corporations, and auditing firms—need to verify its financial status. However, corporate financial records contain highly sensitive information, making it essential to restrict access on the basis of each entity's role. In traditional loan assessments, companies submit financial statements to banks, which then share them with other stakeholders, creating the risk of information leakage. Additionally, companies face a burdensome process of disclosing extensive financial records, resulting in prolonged screening times.

Our proposed system employs SD-JWT and ABE to manage corporate financial information securely and allow each entity to access only the necessary data. First, an Issuer (such as a company’s accounting department or tax authority) issues a VC containing financial statements and tax records. The Holder (company) encrypts these records using ABE, defining decryption policies on the basis of attributes such as "Bank," "Credit Guarantee Corporation," and "Auditing Firm." For example, the bank can decrypt "Revenue, Profit, and Debt Ratio," the credit guarantee corporation can access "Past Credit Ratings and Repayment History," and the auditing firm can verify "Audited Financial Statements Only."

Each Verifier can independently decrypt relevant parts of the VC on the basis of their role and conduct the necessary assessments. The bank evaluates the company’s financial health, the credit guarantee corporation assesses loan risks, and the auditing firm verifies financial accuracy. This mechanism ensures that each institution can perform independent verification while preventing excessive data exposure.





\subsubsection{Use Case 3: Verifiable Credentials for Supply Chain Tracking}
In supply chain management, tracking the movement of goods through multiple stages—such as land transportation, maritime shipping, and customs clearance—requires a secure and transparent system for information sharing. However, different players in the supply chain (e.g., logistics providers, shipping companies, customs authorities) often have varying levels of information access, making it crucial to implement selective data sharing to ensure security and confidentiality.

By using the combination of SD-JWT and ABE, a VC can be attached to a shipment or product that securely encapsulates all relevant details, such as the origin, destination, and compliance certifications. Each piece of information is encrypted and structured as Disclosures within the SD-JWT, while ABE enforces access control on the basis of the attributes of the Verifier.

For example, during the transportation of a shipment:
\begin{itemize}
    \item Land Transportation Companies: A trucking company may need access to basic details such as the shipment ID, destination, and delivery instructions. Their role as a Verifier would allow them to decrypt only these attributes, ensuring they do not access sensitive details like the shipment's total value or regulatory compliance documents.
    \item Shipping Companies: A maritime carrier might require access to additional information, such as the shipment’s weight, dimensions, and hazardous material certifications, but not the final consignee’s details or pricing information.
    \item Customs Authorities: Customs offices often require access to compliance-related data, such as certificates of origin, tax classification, and product descriptions, to process imports or exports. However, they do not need access to internal logistics details like delivery instructions or internal tracking numbers.
\end{itemize}

By integrating SD-JWT and ABE, each Verifier in the supply chain receives access only to the information relevant to their role, improving security while maintaining operational transparency. The shipment's owner (e.g., the manufacturer or supplier) retains full control over the shared data, determining which attributes each Verifier can decrypt.

It minimizes the risk of data breaches by ensuring that sensitive information, such as the value of goods or proprietary product details, is not unnecessarily exposed to unauthorized parties. Studies on blockchain-based supply chains have highlighted the importance of data privacy and selective sharing for improving trust and operational efficiency in global logistics \cite{b17}\cite{b18}.

\section{Feasibility Verification of proposed method}
\subsection{Evaluation}
This study evaluated the feasibility of a mechanism in which SD-JWT Disclosures are encrypted using ABE and selectively decrypted on the basis of the attributes of the Verifier. To evaluate the feasibility of this approach, a virtual machine was deployed on a physical machine running Windows 10 Pro (16GB RAM, Core i7 3.2GHz). The virtual machine was configured with Ubuntu 22.04 (10.74GB RAM, 6 processors), where the SD-JWT creation, encryption, and decryption processes were measured. The SD-JWT module from Open Wallet Foundation was used.
The ABE module was from NTT Social Informatics Laboratories. 
The ABE module was configured to guarantee 128-bit safety.
The number of Disclosures was varied between 5, 10, and 20 to analyze the impact on processing time.

The times required to generate an SD-JWT were 5.0, 4.3, and 6.2ms on average for 5, 10, and 20 Disclosures, respectively. These results indicate that SD-JWT generation time remains relatively constant regardless of the number of Disclosures, demonstrating that this process has minimal computational overhead and is unlikely to be a performance bottleneck even for a large number of Disclosures.

The total processing times on the client side, including SD-JWT creation and encryption, were 150.5, 283.8, and 561.4ms on average for 5, 10, and 20 Disclosures, respectively. This shows a nearly linear increase in encryption processing time as the number of Disclosures increases. Notably, when the number of Disclosures reached 20, the processing time exceeded 500ms, indicating a considerable computational cost associated with encryption.

Regarding decryption processing time, the average times were 137.9, 250.5, and 488.8ms for 5, 10, and 20 Disclosures, respectively. Similar to encryption, decryption time increased proportionally to the number of Disclosures. In the case of 20 Disclosures, the decryption time approached 500ms, suggesting that decryption, like encryption, also incurs a significant computational cost.

Finally, SD-JWT verification time was measured. The results showed an average of 5.0, 7.2, and 10.4ms for 5, 10, and 20 Disclosures, respectively. Unlike encryption and decryption, verification time remained relatively short, regardless of the number of Disclosures. Since SD-JWT verification mainly involves signature validation and hash consistency checks, it does not impose significant computational overhead, making it feasible for practical implementations.

\subsection{Discussion}
The experimental results indicate that SD-JWT generation itself has a low computational cost and does not pose a significant burden in practical applications. However, ABE-based encryption and decryption time increases in proportion to the number of Disclosures, making optimization necessary for certain use cases.

For applications in fields such as healthcare, financial transactions, and supply chain tracking, where strict privacy control is required, the processing time constraints are relatively relaxed, making the proposed method a feasible solution. However, in scenarios requiring real-time authentication, such as IoT device access control, it may be necessary to limit the number of Disclosures or optimize the computational cost to ensure seamless performance. Specifically, when the number of Disclosures exceeds 20, decryption times exceeding 500ms may impact usability, requiring performance tuning.

Another key consideration is the decryption workload on the Verifier side. While this may not be a concern for server-based environments, it could impact mobile devices or resource-constrained environments where computational power is limited. Potential optimizations include offloading decryption processes to cloud-based services or applying more efficient ABE algorithms to reduce computational overhead. Additionally, minimizing the number of attributes assigned to each Disclosure could improve overall performance when defining decryption policies.


\section{Conclusion}
This study proposes a mechanism for encrypting SD-JWT Disclosures using ABE and controlling decryption on the basis of the attributes of the Verifier. SD-JWT is issued by an Issuer to a Holder, who selectively discloses certain Disclosures to the Verifier. However, existing SD-JWT implementations do not provide dynamic access control on the basis of specific conditions, making it difficult to manage decryption flexibly. In the proposed method, the Holder encrypts Disclosures using CP-ABE and assigns decryption policies, allowing information to be selectively disclosed on the basis of the Verifier’s attributes.

To evaluate the feasibility of this approach, a virtual machine was deployed on a physical machine, and processing times for SD-JWT generation, encryption, and decryption were measured. 

These results indicate that SD-JWT generation is a lightweight process and does not impose a significant burden, even with a large number of Disclosures. However, the computational cost of ABE-based encryption and decryption increases proportionally with the number of Disclosures, necessitating optimization for practical applications. The approach is suitable for healthcare applications, financial transactions and supply chain tracking where processing time constraints are relatively flexible. 


\section*{Acknowledgment}
ChatGPT was used to translate the text in this paper into English and to formulate the documents.

\end{document}